\documentclass[twocolumn,prl,showpacs]{revtex4}

\usepackage{epsfig}
\usepackage{graphics}
\usepackage{amssymb,amsmath,amsbsy}
\usepackage{graphicx}
\usepackage{psfrag}
\usepackage[latin2]{inputenc}
\usepackage{bm}

\newcommand{\e}{\mathrm{e}}                  
\newcommand{\ket}[1]{|#1\rangle}             
\newcommand{\bra}[1]{\langle #1|}            
\newcommand{\mt}[1]{\mathrm{#1}}             

\newcommand{\bs}[1]{\boldsymbol{#1}}        

\begin{document}

\title{Ramsey interferometry with ultracold atoms}

\author{D.\ Seidel}
\email{seidel@physik.uni-goe.de}
\author{J.\ G.\ Muga}
\email{jg.muga@ehu.es}
\affiliation{Departamento de Qu\'{\i}mica-F\'{\i}sica, Universidad del
Pa\'{\i}s Vasco, Apartado Postal 644, 48080 Bilbao, Spain}

\begin{abstract}
We examine the passage of ultracold two-level atoms through two separated laser fields for the nonresonant case. We show that implications of the atomic quantized motion change dramatically the behavior of the interference fringes compared to the semiclassical description of this optical Ramsey interferometer. Using two-channel recurrence relations we are able to express the double-laser scattering amplitudes by means of the single-laser ones and to give explicit analytical results. When considering slower and slower atoms, the transmission probability of the system changes considerably from an interference behavior to a regime where scattering resonances prevail.
This may be understood in terms of different families of trajectories that dominate the overall transmission probability in the weak field or in the strong field limit.
\end{abstract}

\pacs{42.50.Ct, 03.65.Nk, 03.75.-b, 39.20.+q}
\maketitle

\paragraph{Introduction.} Atom interferometry based on Ramsey's method with separated fields \cite{Ramsey} is an important tool of modern precision measurements. 
A basic feature of the observed Ramsey interference fringes is that its width is simply the inverse of the time taken by the atoms to cross the intermediate region.
For precision measurement purposes, such as atomic clocks, this implicates the desire for very slow (ultracold) atoms. Experimentally, atomic velocities of the order of $1\,\mt{cm/s}$ are within discussion for space-based atomic clocks \cite{Salomon}.
But, if the kinetic energy becomes comparable with the atom-field interaction energy, one has to take into account the quantized center-of-mass motion of the atom and the well-known semiclassical results have to be corrected. 
This has first been considered for ultracold atoms passing through one \cite{Mazer1} or two \cite{Mazer2} micromaser cavities in resonance with the atomic transition. The nonresonant case has been investigated so far only for one field zone with interesting consequences for the induced emission process inside the cavity \cite{BaMa-PRA-2003}. 

In this letter we study the interference fringes with respect to detuning for ultracold atoms passing through two separated laser fields. Note that for quantized motion the trivial relation to the interferometers operation in the time domain (pulsed fields) does not hold anymore.
For convenience we restrict our analysis to classical fields, but we emphasize the fact that our results can be easily carried over to the case of two quantized fields (mazer physics) by using the appropriate single-cavity scattering coefficients given in Ref.~\cite{BaMa-PRA-2003}. 

\psfrag{|1>}{$\ket{1}$}
\psfrag{|2>}{$\ket{2}$}
\psfrag{x}{$x$}
\psfrag{0}{$0$}
\psfrag{l}{$l$}
\psfrag{l+L}{$l+L$}
\psfrag{2l+L}{$2l+L$}
\psfrag{(0)}{(0)}\psfrag{(1)}{(1)}\psfrag{(2)}{(2)}\psfrag{(3)}{(3)}\psfrag{(4)}{(4)}
\begin{figure}
\centering
\epsfxsize = 8.5cm \epsfbox{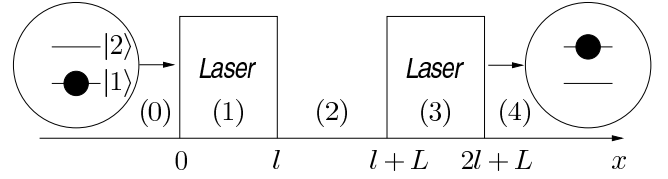}
\caption{Scheme of the optical Ramsey atom interferometer. The regions with constant potential are labeled by $m=0,\dots,4$, respectively.}
\label{fig:setup}
\end{figure}

\paragraph{Model.}
We consider the basic Ramsey setup where a two-level atom in the ground state moves along the $x$ axis on the way to two separated oscillating fields localized between $0$ and $l$ and between $l+L$ and $2l+L$ (Fig.~\ref{fig:setup}). Our one-dimensionless treatment neglects effects connected with transverse momentum components \cite{Borde} and is a good approximation for atoms travelling in narrow waveguides. One asks for the transmission probability of excited atoms, $P_{12}$, in dependence of the laser detuning. In the interaction picture and using the dipole and rotating-wave approximation, the situation may be described by the Hamiltonian
\begin{equation}
H = \frac{\widehat{p}^{\,2}}{2m} - \hbar\Delta \ket{2}\bra{2} + \frac{\hbar}{2}\Omega(\widehat{x}) (\ket{1}\bra{2} + \ket{2}\bra{1}),
\end{equation}
where the first term counts for the kinetic energy of the atom, $\Delta = \omega_L - \omega_{12}$ denotes the detuning between laser frequency and atomic transition frequency and $\Omega(x)$ is the position-dependent Rabi frequency. For the explicit $x$ dependence we assume mesa functions throughout this paper, $\Omega(x) = \Omega$ for $x\in [0,l]$ and $x \in [l+L,2l+L]$ and zero elsewhere. 

In order to derive $P_{12}$ one has to solve the stationary Schr\"odinger equation (SSE), $H\Phi_k = E_k\Phi_k$, with $E_k = \hbar^2 k^2/2m$ and ${\Phi}_k = \phi_k^{(1)}\ket{1} + \phi_k^{(2)}\ket{2}$. This is easy in the semiclassical regime where $E_k \gg \hbar\Omega, \hbar\Delta$ and the center-of-mass motion can be treated independently of the internal dynamics. In this regime and defining the effective Rabi frequency $\Omega' = \sqrt{\Omega^2 + \Delta^2}$ one obtains \cite{Ramsey}
\begin{multline} \label{eq:P12_scl}
P_{12}^\mt{scl} = \frac{4\Omega^2}{\Omega'^2} \sin^2\left(\frac{m\Omega' l}{2\hbar k}\right)  \left[ \cos\left(\frac{m\Delta L}{2\hbar k} \right) \cos\left(\frac{m\Omega' l}{2\hbar k} \right) \right.\\ \left.- \frac{\Delta}{\Omega'} \sin\left(\frac{m\Delta l}{2\hbar k}\right) \sin \left(\frac{m\Omega' l}{2\hbar k} \right)\right]^2.
\end{multline}
However, if the kinetic energy of the atom is comparable with the interaction energy, the semiclassical approach is not valid anymore and a full quantum mechanical solution is required.

\paragraph{Quantum mechanical treatment.} For the given initial value problem of a left incoming plane wave in the ground state the asymptotic solution of the SSE to the right of the laser fields is given by
\begin{equation} \label{eq:Phi_trans}
\Phi_k(x) = \frac{1}{\sqrt{2\pi}} \bigl(T^l_{11} \e^{ikx} \ket{1} + T^l_{12} \e^{iqx} \ket{2} \bigr),~~~x \geq 2l+L,
\end{equation}
where $T^l_{11}$ 
and $T^l_{12}$ 
are the transmission 
amplitudes for the ground and excited state, respectively.

From Eq.~(\ref{eq:Phi_trans}) one sees that after passing the two laser fields, the atom will either be still in the ground state, propagating with momentum $k$ or in the excited state, propagating with momentum $q = \sqrt{k^2+2m\Delta/\hbar}$. In the latter case, the atomic transition $\ket{1} \to \ket{2}$ induced by the laser field is responsible for a change of kinetic energy \cite{BaMa-PRA-2003}. For $\Delta>0$ the kinetic energy of the excited state component has been enhanced by $\hbar\Delta$ whereas for $\Delta <0$ the kinetic energy has been reduced by $\hbar\Delta$ and the laser will slow down the atom. For $\Delta$ smaller then the critical value $\Delta_\mt{cr}=-\hbar k^2/2m$, the excited state component becomes evanescent and its transmission probability vanishes. Thus, the quantum mechanical probability to observe the transmitted atom in the excited state is
\begin{equation}
P_{12} = \frac{q}{k}|T^l_{12}|^2 \quad\mt{for}\quad \Delta > \Delta_\mt{cr}
\end{equation}
and zero elsewhere.

\paragraph{\it Two-channel recurrence relations.}
In the following, we generalize the well-known recurrence relations for one-channel scattering \cite{RoReTe-PRA-1994} to the two-channel case. This allows us to express the double-barrier scattering amplitudes and in particular $T^l_{12}$ in terms of the single-barrier ones.
First we will give the eigenstates of $H$ within a constant potential region and match them at the boundaries. Note that for the nonresonant case it is not possible to find a dressed state basis that diagonalizes $H$ on the whole axis \cite{BaMa-PRA-2003}.
Outside the laser fields (regions $m=\mt{0},\mt{2},\mt{4}$, see Fig.~\ref{fig:setup}) the general solution of the SSE is
\begin{multline}
    \Phi_k^{(m)}(x) = \frac{1}{\sqrt{2\pi}} \left( (A^+_m \e^{i kx} + B^+_m
    \e^{-i kx}) \ket{1}\right.\\ \left. +  (A^-_m \e^{i qx} + B^-_m \e^{-i qx}) \ket{2} \right).
\end{multline}
Within a laser barrier the
(unnormalized) dressed state basis that diagonalizes the Hamiltonian is given by
$\ket{\lambda_\pm} = \ket{1} + 2\lambda_\pm\Omega^{-1} \ket{2}$ where $\lambda_\pm=(-\Delta\pm \Omega')/2$
are the dressed eigenvalues. Thus, the solutions in the interaction regions ($m=\mt{1},\mt{3}$) are of the form
\begin{multline}
    {\Phi}_k^{(m)}(x) = \frac{1}{\sqrt{2\pi}} ( A_m^+
    \ket{\lambda_+} \e^{i k_+ x} + B_m^+
    \ket{\lambda_+} \e^{- i k_+ x}\\  + A_m^-
    \ket{\lambda_-} \e^{i k_- x} + B_m^-
    \ket{\lambda_-} \e^{- i k_- x} )
\end{multline}
with wavenumbers $k_\pm^2 = k^2 -2m\lambda_\pm/\hbar$. For the matching conditions it is convenient to use a two-channel transfer matrix formalism \cite{DaEgHeMu-JPB-2003}. For this we define the column vectors $\bs{v}_m = (A_m^+, B_m^+, A_m^-, B_m^-)^\mt{T}$, $m = 0,\dots,4$,
and the matching matrices $\bs{M}_0(x)$ and $\bs{M}_\mt{b}(x)$, such that the usual matching conditions between the free region and the barrier region at position $x_1$ are given by $\bs{M}_0(x_1) \bs{v}_m = \bs{M}_\mt{b}(x_1) \bs{v}_{m+1}$ and between barrier region and free region at position $x_2$ by $\bs{M}_\mt{b}(x_2) \bs{v}_m = \bs{M}_0(x_2) \bs{v}_{m+1}$. The matrices $\bs{M}_0(x)$ and $\bs{M}_\mt{b}$ are given explicitly in Ref.~\cite{DaEgHeMu-JPB-2003}.
Then the matching conditions between the regions $m=0$ and $m=2$ and between regions $m=0$ and $m=4$ read
\begin{equation} \label{eq:transfer024}
 \bs{v}_0 = \bs{\alpha}\cdot\bs{v}_2 = \bs{\alpha}\tilde{\bs{\alpha}}\cdot \bs{v}_4,
\end{equation}
where $\bs{\alpha} = \bs{N}^{-1}_0(0) \bs{N}(0) \bs{N}^{-1}(l) \bs{N}_0(l)$ and $\tilde{\bs{\alpha}} = \bs{N}^{-1}_0(l+L) \bs{N}(l+L) \bs{N}^{-1}(2l+L) \bs{N}_0(2l+L)$. With Eq.~(\ref{eq:transfer024}), the scattering amplitudes of the single-barrier problem are given in terms of the $\alpha_{ij}$ for any initial value problem. We denote in the following by $r^l_{ij} (t^l_{ij})$ the single-laser reflection (transmission) amplitude for incidence from the left in the $i$th channel and a outgoing plane wave in the $j$th channel, and by $r^r_{ij}$ and $t^r_{ij}$ the corresponding amplitudes for right incidence. The corresponding quantities for the second laser barrier will differ only by a phase factor and they are denoted by a tilde, $\tilde{r}^{l,r}_{ij}, \tilde{t}^{l,r}_{ij}$. One can show that the relation between the 16 scattering amplitudes $r^{l,r}_{ij}, t^{l,r}_{ij}$ and the $\alpha_{ij}$ is invertible, thus $\alpha_{ij} = f(r^{l,r}_{ij}, t^{l,r}_{ij})$ and $\tilde{\alpha}_{ij} = g(\tilde{r}^{l,r}_{ij}, \tilde{t}^{l,r}_{ij})$ with known functions $f$ and $g$. Since according to Eq.~(\ref{eq:transfer024}) the scattering amplitudes of the double-barrier problem are given in terms of $\alpha_{ij}$ and $\tilde{\alpha}_{ij}$, this shows the desired connection between single barrier and double barrier case. In particular, we obtain for our purpose
\begin{eqnarray} \label{eq:T12}
T^l_{12} &=& \Bigl[ t^l_{12}\tilde{t}^l_{22} + t^l_{11}\tilde{t}^l_{12} - (r^r_{12}t^l_{11} - r^r_{11}t^l_{12})(\tilde{r}^l_{21} \tilde{t}^l_{12} - \tilde{r}^l_{11}\tilde{t}^l_{22}) \nonumber\\
    && + (r^r_{22}t^l_{11} - r^r_{21}t^l_{12})(\tilde{r}^l_{22} \tilde{t}^l_{12} - \tilde{r}^l_{12}\tilde{t}^l_{22}) \Bigr] \nonumber\\
    &&\times \Bigl[1 - r^r_{12}\tilde{r}^l_{21} - r^r_{22}\tilde{r}^l_{22} - r^r_{11} \tilde{r}^l_{11} - r^r_{21} \tilde{r}^l_{12}\nonumber\\
    && - (r^r_{12}r^r_{21} - r^r_{22}r^r{11})(\tilde{r}^l_{11} \tilde{r}^l_{22} - \tilde{r}^l_{21} \tilde{r}^l_{12}) \Bigr]^{-1}.
\end{eqnarray}
We will show in the following that a clear physical interpretation of this result can be given within a multiple scattering picture.

\begin{figure}
\centering
\epsfxsize = 8.4cm \epsfbox{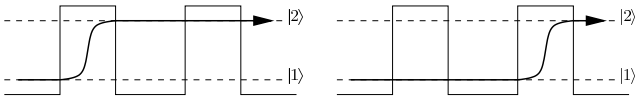}
\caption{Leading order contribution to the multiple scattering paths for the direct scattering regime ($E_k \gg \hbar\Omega$).}
\label{fig:path1}
\end{figure}

\paragraph{\it Direct scattering regime ($E_k \gg \hbar\Omega$).}
If the kinetic energy is larger than the Rabi energy the scattering process will be dominated by transmission through both laser barriers. Thus, one has $t^{l,r}_{ij} \gg r^{l,r}_{ij}$ and expanding the denominator of Eq.~(\ref{eq:T12}) yields
\begin{multline} 
    T^l_{12} = \bigl[t^l_{11}\tilde{t}^l_{12} + t^l_{12}\tilde{t}^l_{22}\bigr] + \bigl[
    t^l_{12} (\tilde{r}^l_{21}r^r_{11} + \tilde{r}^l_{22}r^r_{21})\tilde{t}^l_{12}  \\
  + t^l_{11}(\tilde{r}^l_{11} r^r_{12} + \tilde{r}^l_{12}r^r_{22}) \tilde{t}^l_{22}
    + t^l_{12} (\tilde{r}^l_{21}r^r_{12} + \tilde{r}^l_{22}r^r_{22})\tilde{t}^l_{22}\\
   + t^l_{11}(\tilde{r}^l_{11} r^r_{11} + \tilde{r}^l_{12}r^r_{21}) \tilde{t}^l_{12} \bigr] + \cdots.
\end{multline}
The two terms in the first bracket of this expansion describe the direct scattering process (Fig.~\ref{fig:path1}) whereas the second bracket contains all possible paths to first order of multiple scatterings including two reflections. 

Keeping only the two direct scattering paths and using $\tilde{t}^l_{22} = t^l_{22}$ and $\tilde{t}^l_{12} = t^l_{12}\exp(i(k-q)(l+L))$ leads to
\begin{equation} \label{eq:P12_hot}
 P_{12} = \frac{q}{k}|t^l_{12}|^2 \left|t^l_{22} + t^l_{11} \e^{i(k-q)(l+L)}\right|^2.
\end{equation}
This expression describes interferences between atoms which pass the first laser in the ground state and are excited in the second laser and atoms which are excited in the first laser and pass the second one in the excited state (Fig.~\ref{fig:path1}). However, we emphasize the fact that Eq.~(\ref{eq:P12_hot}) goes beyond the validity of the semiclassical expression (\ref{eq:P12_scl}) since $E_k$ has not to be large with respect to $\hbar\Delta$
and the overall transmission probability might be smaller than one.

The explicit expressions for the single-barrier amplitudes are $t^l_{11} = -\alpha_{33}/d_\alpha$, $t^l_{12} = \alpha_{31}/d_\alpha$ and $t^l_{22} = -\alpha_{11}/d_\alpha$ 
where $d_\alpha = \alpha_{13}\alpha_{31} - \alpha_{11}\alpha_{33}$ and 
\begin{eqnarray}
\alpha_{11} &=& \e^{ikl} [\lambda_+d_-(k,k)-\lambda_-d_+(k,k)]/\Omega',\nonumber\\
\alpha_{33} &=& \alpha_{11}(k\to q), \nonumber\\ \label{eq:alphas1}
\alpha_{31} &=& \textstyle\frac{1}{4}\Omega\e^{ikl}\Bigl[d_+(k,q) - d_-(k,q) \\ && 
~~~~~~~~ + \frac{k}{q}\Bigl(d_+(q,k)-d_-(q,k)\Bigr)\Bigr] \Big/\Omega', \nonumber \\
\alpha_{13} &=& \alpha_{31}(k \leftrightarrow q), \nonumber \\
d_\pm(k_1,k_2) &=&\cos(k_\pm l) - \sigma_\pm(k_1,k_2)\sin(k_\pm l), \nonumber \\
\sigma_\pm(k_1,k_2) &=& i(k_1/k_\pm + k_\pm/k_2)/2. \nonumber
\end{eqnarray}
The given replacement rules apply only to the exponents and arguments and not to the indirect dependencies within the $k_\pm$. 
Expanding $P_{12}$ with respect to $k\to\infty$ gives back the semiclassical result (\ref{eq:P12_scl}), as expected.

\begin{figure}
\centering
\epsfxsize=8.5cm  \epsfbox{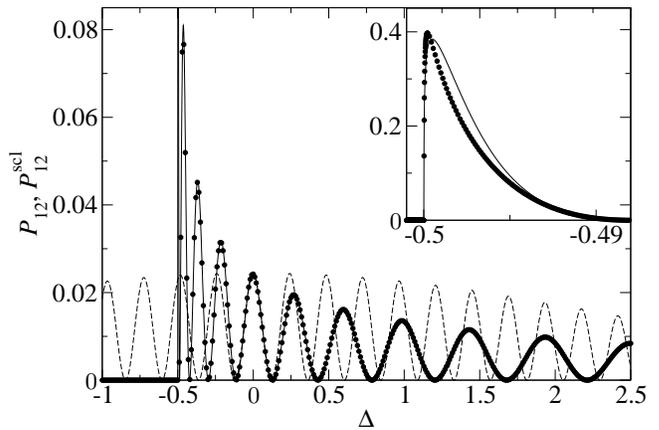}
\caption{The detection probability $P_{12}$ with respect to $\Delta$ in the direct scattering regime ($E_k \gg \hbar\Omega$). Comparison of exact result obtained with Eq.~(\ref{eq:T12}) (solid line), direct scattering approximation (\ref{eq:P12_hot}) (circles) and semiclassical result $P_{12}^\mt{scl}$ (dashed line) for $k=1$, $\Omega = \pi/20$, $l=1$,  $L=25$. The inset shows a closeup of the peak close to $\Delta_\mt{cr}$. For plots we use dimensionless units with $\hbar = m = 1$.}
\label{fig:scl1}
\end{figure}

Figure \ref{fig:scl1} illustrates the the behavior of $P_{12}$ as a function of $\Delta$. For comparison we plotted the exact result obtained by means of Eqs.~(\ref{eq:T12}) and the semiclassical expression $P_{12}^\mt{scl}$.
For $E_k/\hbar\Omega \approx 3.2$ the agreement between $P_{12}$ and the exact result is very good and becomes even better for larger velocities. Thus, the scattering process is dominated by the two paths shown in Fig.~\ref{fig:path1}. Nevertheless, one sees that the interference pattern can not be understood within the semiclassical approximation for $\hbar\Delta \gtrsim E_k$. For $\Delta > 0$, the zeros of the interference pattern become dispersed with respect to $P_{12}^\mt{scl}$ whereas for $\Delta_\mt{cr} < \Delta < 0$ the fringes become closer and narrower up to extremely narrow peaks close to $\Delta_\mt{cr}$ (see inset of Fig.~\ref{fig:scl1}). Physically, this is related to the discussion below Eq.~(\ref{eq:Phi_trans}). For negative detuning and close to the critical value the absorbing process will slow down the atom and the effective crossing time scale $t_\mt{eff} = mL/\hbar q$ will be much larger than the semiclassical crossing time scale $t_\mt{scl} = mL/\hbar k$. On the other hand, for positive detuning the atom is speed up and $t_\mt{eff} < t_\mt{scl}$, leading to a broader interference pattern. This suggests to use the negative detuning regime close to $\Delta_\mt{cr}$ for metrology purposes. Note that for current experiments the region close to $\Delta=0$ where the semiclassical condition is well satisfied contains much more fringes than shown in Fig.~(\ref{fig:scl1}).

\begin{figure}
\centering
\epsfxsize = 8.4cm \epsfbox{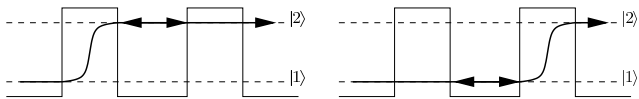}
\caption{Leading order contributions to the multiple scattering paths for $E_k \ll \hbar\Omega$.}
\label{fig:path2}
\end{figure}

\paragraph{\it Ultracold regime ($E_k \ll \hbar\Omega$).}
Let us now consider the opposite regime, where the kinetic energy of the atom is much smaller than the Rabi energy. In this case one has $r^{l,r}_{12},r^{l,r}_{21},t^{l,r}_{ij} \ll r^{l,r}_{11}, r^{l,r}_{22}$. We checked explicitly that the leading order of the small scattering amplitudes is $(E_k/\hbar\Omega)^{1/2}$, respectively, whereas the leading order of $r^{l,r}_{11}$ and $r^{l,r}_{22}$ is 1. Thus, an expansion of $T^l_{12}$ in a series with respect to powers of the small amplitudes yields
\begin{multline} \label{eq:T12_ultra}
\!\!\!\!T^l_{12} = \frac{t^l_{12} \tilde{t}^l_{22}}{1- r^r_{22} \tilde{r}^l_{22}} +  \frac{t^l_{11} \tilde{t}^l_{12}}{1- r^r_{11} \tilde{r}^l_{11}}
    + \frac{t^l_{11} \tilde{t}^l_{22} (r^r_{12} \tilde{r}^l_{11} + r^r_{22} \tilde{r}^l_{12})}{(1- r^r_{11} \tilde{r}^l_{11})(1- r^r_{22} \tilde{r}^l_{22})} \\
 +  \frac{t^l_{12} \tilde{t}^l_{12} (r^r_{11} \tilde{r}^l_{21} + r^r_{22} \tilde{r}^l_{22})}{(1- r^r_{11} \tilde{r}^l_{11})(1- r^r_{22} \tilde{r}^l_{22})}+ \dots,
\end{multline}
where the first two terms contain second powers of the small scattering amplitudes, the next given terms contain third powers and so on. Again, these terms can be given a clear physical interpretation in terms of multiple scatterings. The first two terms correspond to two families of paths, the first of which describes atoms being excited in the first laser and cross the second laser in the excited state after multiple internal reflections, whereas the second family corresponds to atoms crossing the first barrier in the ground state, undergo multiple internal reflections in the ground state and are finally excited in the second laser (Fig.~\ref{fig:path2}).
Similarly, the following terms account for all possible paths with exactly one internal excitation by means of reflection. Keeping only the leading order of Eq.~(\ref{eq:T12_ultra}) and using $r^r_{22} = r^l_{22} \exp(-2iql)$ and $r^r_{11} = r^l_{11}\exp(-2ikl)$, we obtain for the probability to observe an outgoing atom in the excited state
\begin{equation} \label{eq:P12_ultra}
 P_{12} = \frac{q}{k} \left| \frac{t^l_{12} t^l_{22}}{1- (r^l_{22})^2 \e^{2iqL}} +  \frac{t^l_{11} t^l_{12}\e^{i(k-q)(l+L)}}{1- (r^l_{11})^2 \e^{2ikL}} \right|^2.
\end{equation}
Again, we give explicit analytical results by writing the scattering coefficients in terms of the $\alpha_{ij}$. We find $r^l_{11} = (\alpha_{23} \alpha_{31} - \alpha_{21} \alpha_{33})/d_\alpha$ and $r^l_{22} = (\alpha_{41} \alpha_{13} - \alpha_{43} \alpha_{11})/d_\alpha$
where additionally to Eqs.~(\ref{eq:alphas1}) one has
\begin{eqnarray}
\alpha_{41} &=& \alpha_{31}(q \to -q), \nonumber\\
\alpha_{23} &=& \alpha_{13}(k \to -k), \nonumber\\
\alpha_{21} &=& \e^{ikl}[\lambda_-\sigma_+(k,-k) - \lambda_+\sigma_-(k,-k)]/\Omega', \nonumber\\
\alpha_{43} &=& \alpha_{21}(k \to q).
\end{eqnarray}
For the resonant case, $\Delta=0$, one has $t^l_{11}=t^l_{22}=(\tau_+ +\tau_-)/2$, $t^l_{12}=(\tau_+ -\tau_-)/2$, $r^l_{11}=r^l_{22}=(\rho_+ +\rho_-)/2$ and $P_{12}$ can be written in terms of the one-channel double barrier and double well amplitudes $\tau_\pm$ and $\rho_\pm$ \cite{Mazer2}.

\begin{figure}[t]
\centering
\epsfxsize = 8.5cm \epsfbox{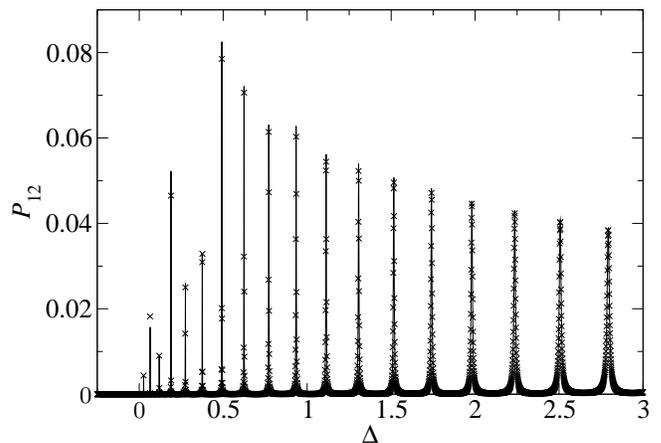}
\caption{The detection probability $P_{12}$ with respect to $\Delta$ in the ultracold  regime ($E_k \ll \hbar\Omega$). Comparison of the exact result obtained with Eq.~(\ref{eq:T12}) (solid line) and the approximation (\ref{eq:P12_hot}) (crosses) for $k=0.1$, $\Omega = 15\pi$, $l=1.2$,  $L=25$.}
\label{fig:ultra}
\end{figure}

Eq.~(\ref{eq:P12_ultra}) may be understood as a coherent sum of two Fabry-Perot-like terms. Both of them exhibit resonances with respect to $k$, but only the first one shows resonances with respect to $\Delta$. The reason is that only the excited state sums up a $\Delta$-dependent phase while crossing the intermediate region.
A typical picture is given in Fig.~\ref{fig:ultra}, where $P_{12}$ is plotted for $E_k/\hbar\Omega \approx 10^{-4}$. With this ratio, it can be seen that Eq.~(\ref{eq:P12_ultra}) is in excellent agreement with the exact result. 

The position of the resonance peaks can be estimated by considering the leading order of the reflection amplitudes in the denominators of Eq.~(\ref{eq:P12_ultra}), $r^l_{11} \approx r^l_{22} \approx -1$. Then the peaks with respect to $\Delta$ are given by the zeros of $1-\exp(2iqL)$ which leads to $\Delta_n \approx [(n\pi/L)^2 -k^2]/2,\qquad n= 1,2,3,\dots$, provided that $k\neq n\pi/L$. The resonances of Fig.~\ref{fig:ultra} are in the vicinity of these values. Note that the result in the ultracold case is no more an interference between two paths but the resonance behavior of the family of paths shown in the left part of Fig.~\ref{fig:path2}.

\paragraph{Conclusion.} We studied analytically the behavior of Ramsey fringes beyond the semiclassical approximation. By means of two-channel recurrence relations we were able to identify the dominant contributions to the scattering process in the direct scattering regime and in the ultracold regime. We have shown that for $E_k \gg \hbar\Omega$ the interference fringes for negative detuning are narrower than semiclassical theory predicts. For ultracold atoms, $E_k \ll \hbar\Omega$, interference is completely suppressed and the transmission probability is dominated by scattering resonances.
Clearly the current work can be adopted to quantized fields to describe the passage of ultracold atoms through two microwave high-Q cavities. This would be highly interesting in view of the recent results obtained in Ref.~\cite{AgPaSc-PRA-2003}.

This work has been supported by Ministerio de Edu\-ca\-ci\'on y Ciencia (BFM2003-01003) and UPV-EHU (00039.310-15968/2004). D.S.\ acknowledges financial support by the DAAD. 

\end{document}